\documentclass{article}%
\usepackage{amssymb}
\usepackage{amsmath}
\usepackage{amsfonts}
\usepackage{graphicx}%
\providecommand{\U}[1]{\protect\rule{.1in}{.1in}}
\newtheorem{theorem}{Theorem}[section]

\newtheorem{corollary}[theorem]{Corollary}

\newtheorem{lemma}[theorem]{Lemma}

\newtheorem{problem}[theorem]{Problem}
\newtheorem{proposition}[theorem]{Proposition}
\newtheorem{remark}[theorem]{Remark}

\newenvironment{proof}[1][Proof]{\noindent\textbf{#1.} }{\ \rule{0.5em}{0.5em}}
\begin{document}

\title{A Set and Collection Lemma}
\author{Vadim E. Levit\\Ariel University Center of Samaria, Israel\\levitv@ariel.ac.il
\and Eugen Mandrescu\\Holon Institute of Technology, Israel\\eugen\_m@hit.ac.il}
\date{}
\maketitle

\begin{abstract}
A set $S\subseteq V(G)$ is \textit{independent} if no two vertices from $S$
are adjacent. Let $\alpha\left(  G\right)  $ stand for the cardinality of a
largest independent set.

In this paper we prove that if $\Lambda$ is a n\emph{on-em}pty collection of
maximum independent sets of a graph $G$, and $S$ is an independent set, then

\begin{itemize}
\item there is a matching from $S-\cap\Lambda$ into $\cup\Lambda-S$, and

\item $\left\vert S\right\vert +\alpha(G)\leq\left\vert \cap\Lambda\cap
S\right\vert +\left\vert \cup\Lambda\cup S\right\vert $.

\end{itemize}

Based on these findings we provide alternative proofs for a number of
well-known lemmata, as the \textquotedblleft\textit{Maximum Stable Set
Lemma}\textquotedblright\ due to Claude Berge and the \textquotedblleft%
\textit{Clique Collection Lemma}\textquotedblright\ due to Andr\'{a}s Hajnal.

\textbf{Keywords:} matching, independent set, stable set, core, corona, clique

\end{abstract}

\section{Introduction}

Throughout this paper $G=(V,E)$ is a simple (i.e., a finite, undirected,
loopless and without multiple edges) graph with vertex set $V=V(G)$ and edge
set $E=E(G)$. If $X\subseteq V$, then $G[X]$ is the subgraph of $G$ spanned by
$X$. By $G-W$ we mean the subgraph $G[V-W]$, if $W\subseteq V(G)$, and we use
$G-w$, whenever $W$ $=\{w\}$.

The \textit{neighborhood} of a vertex $v\in V$ is the set $N(v)=\{w:w\in V$
\ \textit{and} $vw\in E\}$, while the \textit{neighborhood} of $A\subseteq V$
is $N(A)=N_{G}(A)=\{v\in V:N(v)\cap A\neq\emptyset\}$. By $\overline{G}$ we
denote the complement of $G$.

A set $S\subseteq V(G)$ is \textit{independent} (\textit{stable}) if no two
vertices from $S$ are adjacent, and by $\mathrm{Ind}(G)$ we mean the set of
all the independent sets of $G$. An independent set of maximum cardinality
will be referred to as a \textit{maximum independent set} of $G$, and the
\textit{independence number }of $G$ is $\alpha(G)=\max\{\left\vert
S\right\vert :S\in\mathrm{Ind}(G)\}$.

A matching (i.e., a set of non-incident edges of $G$) of maximum cardinality
$\mu(G)$ is a \textit{maximum matching}. If $\alpha(G)+\mu(G)=\left\vert
V\left(  G\right)  \right\vert $, then $G$ is called a
\textit{K\"{o}nig-Egerv\'{a}ry graph }\cite{Dem1979, Ster1979}.

Let $\Omega(G)$ denote the family of all maximum independent sets of $G$ and
\begin{align*}
\mathrm{core}(G)  &  =\cap\{S:S\in\Omega(G)\}\text{ \cite{LevMan2002a}, while
}\\
\mathrm{corona}(G)  &  =\cup\{S:S\in\Omega(G)\}\text{ \cite{BorosGolLev2002}}.
\end{align*}

A set $A\subseteq V(G)$ is a \textit{clique} in $G$ if $A$ is independent in
$\overline{G}$, and $\omega\left(  G\right)  =\alpha\left(  \overline
{G}\right)  $.

In this paper we introduce the \textquotedblleft\textit{Set and Collection
Lemma}\textquotedblright. It is both a generalization and strengthening of a
number of elegant observations including the \textquotedblleft\textit{Maximum
Stable Set Lemma}\textquotedblright\ due to Berge and the \textquotedblleft%
\textit{Clique Collection Lemma}\textquotedblright\ due to Hajnal.

\section{Results}

It is clear that the statement \textquotedblleft\textit{there exists a
matching from a set} $A$ \textit{into a set} $B$\textquotedblright\ is
stronger than just saying that $\left\vert A\right\vert \leq\left\vert
B\right\vert $. The \textquotedblleft\textit{Matching Lemma}\textquotedblright%
\ offers both a powerful tool validating existence of matchings and its most
important corresponding inequalities, emphasized in the \textquotedblleft%
\textit{Set and Collection Lemma}\textquotedblright\ and its corollaries.

\begin{lemma}
[Matching Lemma]\label{MatchLem}Let $S\in\mathrm{Ind}(G),X\in\Lambda
\subseteq\Omega(G),\left\vert \Lambda\right\vert \geq1$. Then the following
assertions are true:

\emph{(i)} there exists a matching from $S-\cap\Lambda$ into $\cup\Lambda-S$;

\emph{(ii)} there is a matching from $S-X$ into $X-S$;

\emph{(iii) }there exists a matching from $S\cap X-\cap\Lambda$ into
$\cup\Lambda-\left(  X\cup S\right)  $.
\end{lemma}

\begin{proof}
Let $B_{1}=\cap\Lambda$ and $B_{2}=\cup\Lambda$.

\emph{(i) }In order to prove that there is a matching from $S-B_{1}$ into
$B_{2}-S$, we use Hall's Theorem, i.e., we show that for every $A\subseteq
S-B_{1}$ we must have
\[
\left\vert A\right\vert \leq\left\vert N\left(  A\right)  \cap B_{2}%
\right\vert =\left\vert N\left(  A\right)  \cap\left(  B_{2}-S\right)
\right\vert \text{.}%
\]

Assume, in a way of contradiction, that Hall's condition is not satisfied. Let
us choose a minimal subset $\tilde{A}\subseteq S-B_{1}$, for which $\left\vert
\tilde{A}\right\vert >\left\vert N\left(  \tilde{A}\right)  \cap
B_{2}\right\vert $.

There exists some $W\in\Lambda$ such that $\tilde{A}\nsubseteq W$, because
$\tilde{A}\subseteq S-B_{1}$. Further, the inequality $\left\vert \tilde
{A}\cap W\right\vert <\left\vert \tilde{A}\right\vert $ and the inclusion%
\[
N(\tilde{A}\cap W)\cap B_{2}\subseteq N(\tilde{A})\cap B_{2}-W
\]
imply
\[
\left\vert \tilde{A}\cap W\right\vert \leq\left\vert N(\tilde{A}\cap W)\cap
B_{2}\right\vert \leq\left\vert N(\tilde{A})\cap B_{2}-W\right\vert ,
\]
because we have selected $\tilde{A}$ as a minimal subset satisfying
$\left\vert \tilde{A}\right\vert >\left\vert N\left(  \tilde{A}\right)  \cap
B_{2}\right\vert $. Therefore,%
\[
\left\vert \tilde{A}\cap W\right\vert +\left\vert \tilde{A}-W\right\vert
=\left\vert \tilde{A}\right\vert >\left\vert N(\tilde{A})\cap B_{2}\right\vert
=\left\vert N(\tilde{A})\cap B_{2}-W\right\vert +\left\vert N(\tilde{A})\cap
W\right\vert .
\]
Consequently, since $\left\vert \tilde{A}\cap W\right\vert \leq\left\vert
N(\tilde{A})\cap B_{2}-W\right\vert $, we infer that $\left\vert \tilde
{A}-W\right\vert >\left\vert N(\tilde{A})\cap W\right\vert $. Thus
\[
\tilde{A}\cup\left(  W-N(\tilde{A})\right)  =W\cup\left(  \tilde{A}-W\right)
-\left(  N(\tilde{A})\cap W\right)
\]
is an independent set of size greater than $\left\vert W\right\vert
=\alpha\left(  G\right)  $, which is a contradiction that proves the claim.

\emph{(ii)} It follows from part \emph{(i) }for $\Lambda=\left\{  X\right\}  $.

\emph{(iii)} By part \emph{(i)}, there exists a matching from $S-\cap\Lambda$
into $\cup\Lambda-S$, while by part \emph{(ii)},\emph{ }there is a matching
from $S-X$ into $X-S$. Since $X$ is independent, there are no edges between
\[
\left(  S-B_{1}\right)  -\left(  S-X\right)  =\left(  S\cap X\right)
-B_{1}\text{ and }X-S.
\]
Therefore, there exists a matching
\[
\text{from }\left(  S\cap X\right)  -B_{1}\text{ into }\left(  B_{2}-S\right)
-\left(  X-S\right)  =B_{2}-\left(  X\cup S\right)  ,
\]
as claimed.
\end{proof}

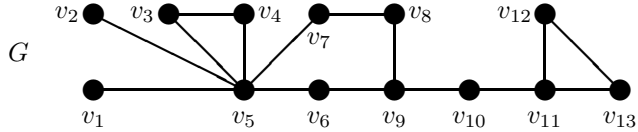
\begin{figure}[h]
\setlength{\unitlength}{1cm}\begin{picture}(5,1.9)\thicklines
\multiput(5,0.5)(1,0){6}{\circle*{0.29}}
\multiput(4,1.5)(1,0){4}{\circle*{0.29}}
\multiput(3,0.5)(0,1){2}{\circle*{0.29}}
\put(9,1.5){\circle*{0.29}}
\put(3,0.5){\line(1,0){7}}
\put(3,1.5){\line(2,-1){2}}
\put(4,1.5){\line(1,-1){1}}
\put(4,1.5){\line(1,0){1}}
\put(5,0.5){\line(0,1){1}}
\put(5,0.5){\line(1,1){1}}
\put(6,1.5){\line(1,0){1}}
\put(7,0.5){\line(0,1){1}}
\put(9,0.5){\line(0,1){1}}
\put(9,1.5){\line(1,-1){1}}
\put(3,0.1){\makebox(0,0){$v_{1}$}}
\put(2.65,1.5){\makebox(0,0){$v_{2}$}}
\put(3.65,1.5){\makebox(0,0){$v_{3}$}}
\put(5.35,1.5){\makebox(0,0){$v_{4}$}}
\put(5,0.1){\makebox(0,0){$v_{5}$}}
\put(6,0.1){\makebox(0,0){$v_{6}$}}
\put(6,1.15){\makebox(0,0){$v_{7}$}}
\put(7,0.1){\makebox(0,0){$v_{9}$}}
\put(7.35,1.5){\makebox(0,0){$v_{8}$}}
\put(8.6,1.5){\makebox(0,0){$v_{12}$}}
\put(8,0.1){\makebox(0,0){$v_{10}$}}
\put(9,0.1){\makebox(0,0){$v_{11}$}}
\put(10,0.1){\makebox(0,0){$v_{13}$}}
\put(2,1){\makebox(0,0){$G$}}
\end{picture}\caption{\textrm{core}$(G)=\{v_{1},v_{2},v_{10}\}$ is not a
critical set.}%
\label{fig51}%
\end{figure}

For example, let us consider the graph $G$ from Figure \ref{fig51} and
$S=\left\{  v_{1},v_{4},v_{7}\right\}  \in\mathrm{Ind}(G)$, $\Lambda=\left\{
S_{1},S_{2}\right\}  $, where $S_{1}=\left\{  v_{1},v_{2},v_{3},v_{6}%
,v_{8},v_{10},v_{12}\right\}  $ and $S_{2}=\left\{  v_{1},v_{2},v_{4}%
,v_{6},v_{7},v_{10},v_{13}\right\}  $. Then, there is a matching from
$S-\cap\Lambda=\left\{  v_{4},v_{7}\right\}  $ into $\cup\Lambda-S=$ $\left\{
v_{2},v_{3},v_{6},v_{8},v_{10},v_{12},v_{13}\right\}  $, namely, $M=\left\{
v_{3}v_{4},v_{7}v_{8}\right\}  $. In addition, we have
\[
10=3+7=\left\vert S\right\vert +\alpha\left(  G\right)  \leq\left\vert
\cap\Lambda\cap S\right\vert +\left\vert \cup\Lambda\cup S\right\vert
=1+10=11.
\]

The assertions of Matching Lemma may be false, if the family $\Lambda$ is not
included in $\Omega\left(  G\right)  $. For instance, if $S=\left\{
v_{1},v_{2},v_{4},v_{7},v_{9},v_{12}\right\}  \in\mathrm{Ind}(G)$,
$\Lambda=\left\{  S_{1},S_{2}\right\}  $, where $S_{1}=\left\{  v_{2}%
,v_{3},v_{7}\right\}  $ and $S_{2}=\left\{  v_{1},v_{2},v_{4},v_{6}%
,v_{7},v_{10},v_{12}\right\}  $, then, there is no matching from
$S-\cap\Lambda=\left\{  v_{1},v_{4},v_{9},v_{12}\right\}  $ into $\cup
\Lambda-S=$ $\left\{  v_{3},v_{6},v_{10}\right\}  $. In addition, we see that
\[
12=2\cdot\left\vert S\right\vert \nleqslant\left\vert \cap\Lambda\cap
S\right\vert +\left\vert \cup\Lambda\cup S\right\vert =2+9=11.
\]

\begin{lemma}
[Set and Collection Lemma]If $S\in\mathrm{Ind}(G)$ and $\Lambda\subseteq
\Omega(G),\left\vert \Lambda\right\vert \geq1$, then
\[
\left\vert S\right\vert +\alpha(G)\leq\left\vert \cap\Lambda\cap S\right\vert
+\left\vert \cup\Lambda\cup S\right\vert .
\]

\end{lemma}

\begin{proof}
Let $X\in\Lambda$. By Matching Lemma \emph{(iii)}, there is a matching from
$S\cap X-\cap\Lambda$ into $\cup\Lambda-\left(  X\cup S\right)  $. Hence we
infer that
\begin{gather*}
\left\vert S\cap X\right\vert -\left\vert \cap\Lambda\cap S\right\vert
=\left\vert S\cap X\right\vert -\left\vert \cap\Lambda\cap S\cap X\right\vert
=\\
=\left\vert S\cap X-\cap\Lambda\right\vert \leq\left\vert \cup\Lambda-\left(
X\cup S\right)  \right\vert =\\
=\left\vert \cup\Lambda\cup\left(  X\cup S\right)  \right\vert -\left\vert
X\cup S\right\vert =\left\vert \cup\Lambda\cup S\right\vert -\left\vert X\cup
S\right\vert .
\end{gather*}
Therefore, we obtain that
\[
\left\vert S\cap X\right\vert -\left\vert \cap\Lambda\cap S\right\vert
\leq\left\vert \cup\Lambda\cup S\right\vert -\left\vert X\cup S\right\vert ,
\]
which implies%
\[
\left\vert S\right\vert +\alpha\left(  G\right)  =\left\vert S\right\vert
+\left\vert X\right\vert =\left\vert S\cap X\right\vert +\left\vert X\cup
S\right\vert \leq\left\vert \cap\Lambda\cap S\right\vert +\left\vert
\cup\Lambda\cup S\right\vert ,
\]
as claimed.
\end{proof}

\begin{corollary}
\label{cor3}If $\Lambda\subseteq\Omega(G),\left\vert \Lambda\right\vert \geq
1$, then $2\cdot\alpha(G)\leq\left\vert \cap\Lambda\right\vert +\left\vert
\cup\Lambda\right\vert $.
\end{corollary}

\begin{proof}
Let $S\in\Lambda$. By Set and Collection Lemma, we get\emph{ }that%
\[
2\cdot\alpha\left(  G\right)  =\left\vert S\right\vert +\alpha\left(
G\right)  \leq\left\vert \cap\Lambda\cap S\right\vert +\left\vert \cup
\Lambda\cup S\right\vert =\left\vert \cap\Lambda\right\vert +\left\vert
\cup\Lambda\right\vert ,
\]
as required.
\end{proof}

If $\Lambda=\Omega(G)$, then Corollary \ref{cor3} gives the following.

\begin{corollary}
\label{cor2}For every graph $G$, it is true that
\[
2\cdot\alpha(G)\leq\left\vert \mathrm{core}(G)\right\vert +\left\vert
\mathrm{corona}(G)\right\vert .
\]

\end{corollary}

It is clear that%
\[
\left\vert \mathrm{core}(G)\right\vert +\left\vert \mathrm{corona}%
(G)\right\vert \leq\alpha\left(  G\right)  +\left\vert V\left(  G\right)
\right\vert .
\]
\begin{figure}[h]
\setlength{\unitlength}{1cm} \begin{picture}(5,1.2)\thicklines
\multiput(4,1)(1,0){2}{\circle*{0.29}}
\multiput(4,0)(1,0){6}{\circle*{0.29}}
\multiput(7,1)(1,0){2}{\circle*{0.29}}
\put(4,1){\line(1,0){1}}
\put(4,0){\line(0,1){1}}
\put(4,0){\line(1,0){5}}
\put(5,1){\line(1,-1){1}}
\put(6,0){\line(1,1){1}}
\put(8,0){\line(0,1){1}}
\put(7,0){\line(0,1){1}}
\put(3.6,0){\makebox(0,0){$v_{1}$}}
\put(3.6,1){\makebox(0,0){$v_{2}$}}
\put(5.35,1){\makebox(0,0){$v_{3}$}}
\put(5,0.3){\makebox(0,0){$v_{4}$}}
\put(6,0.35){\makebox(0,0){$v_{5}$}}
\put(7.35,1){\makebox(0,0){$v_{6}$}}
\put(7.3,0.3){\makebox(0,0){$v_{7}$}}
\put(8.35,1){\makebox(0,0){$v_{8}$}}
\put(8.3,0.3){\makebox(0,0){$v_{9}$}}
\put(9.45,0){\makebox(0,0){$v_{10}$}}
\put(2.5,0.5){\makebox(0,0){$G$}}
\end{picture}\caption{The graph $G$ has $\mathrm{core}(G)=\left\{
v_{8},v_{10}\right\}  $.}%
\label{Fig 12}%
\end{figure}

The graph $G$ from Figure \ref{Fig 12} has $V\left(  G\right)
=\mathrm{corona}(G)\cup N\left(  \mathrm{core}(G)\right)  \cup\left\{
v_{5}\right\}  $.

\begin{proposition}
\label{prop2}If $G=\left(  V,E\right)  $ is a graph with a non-empty edge set,
then
\[
\left\vert \mathrm{core}(G)\right\vert +\left\vert \mathrm{corona}%
(G)\right\vert \leq\alpha\left(  G\right)  +\left\vert V\right\vert -1.
\]

\end{proposition}

\begin{proof}
Notice that for every $S\in\Omega\left(  G\right)  $, we have $\mathrm{core}%
(G)\subseteq S\subseteq\mathrm{corona}(G)\subseteq V$, which implies
$\mathrm{corona}(G)-S\subseteq\mathrm{corona}(G)-\mathrm{core}(G)\subseteq
V-\mathrm{core}(G)$.

Assume, to the contrary, that
\[
\left\vert \mathrm{core}(G)\right\vert +\left\vert \mathrm{corona}%
(G)\right\vert \geq\alpha\left(  G\right)  +\left\vert V\right\vert .
\]
Hence we infer that
\[
\left\vert \mathrm{corona}(G)\right\vert -\alpha\left(  G\right)
\geq\left\vert V\right\vert -\left\vert \mathrm{core}(G)\right\vert ,
\]
i.e.,
\[
\left\vert \mathrm{corona}(G)-S\right\vert \geq\left\vert V-\mathrm{core}%
(G)\right\vert .
\]
Since $\mathrm{corona}(G)-S\subseteq V-\mathrm{core}(G)$, we get that
$V=\mathrm{corona}(G)$ and $\mathrm{core}(G)=S$. It follows that $N\left(
\mathrm{core}(G)\right)  =\emptyset$, since $\mathrm{corona}(G)\cap N\left(
\mathrm{core}(G)\right)  =\emptyset$.

On the other hand, $G$ must have $N\left(  \mathrm{core}(G)\right)
\neq\emptyset$, because $G$ has a non-empty edge set and $\mathrm{core}%
(G)=S\neq\emptyset$.

This contradiction proves that the inequality
\[
\left\vert \mathrm{core}(G)\right\vert +\left\vert \mathrm{corona}%
(G)\right\vert \leq\alpha\left(  G\right)  +\left\vert V\right\vert -1
\]
is true.
\end{proof}

\begin{remark}
The complete bipartite $K_{1,n-1}$ satisfies $\alpha\left(  K_{1,n-1}\right)
=n-1$, and hence
\[
\left\vert \mathrm{core}(K_{1,n-1})\right\vert +\left\vert \mathrm{corona}%
(K_{1,n-1})\right\vert =2\left(  n-1\right)  =\alpha\left(  G\right)
+\left\vert V\left(  K_{1,n-1}\right)  \right\vert -1.
\]
In other words, the bound in Proposition \ref{prop2} is tight.
\end{remark}

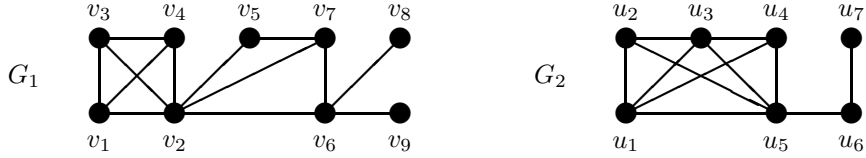
\begin{figure}[h]
\setlength{\unitlength}{1cm}\begin{picture}(5,1.95)\thicklines
\multiput(2,0.5)(1,0){2}{\circle*{0.29}}
\multiput(5,0.5)(1,0){2}{\circle*{0.29}}
\multiput(2,1.5)(1,0){5}{\circle*{0.29}}
\put(2,0.5){\line(1,0){4}}
\put(2,0.5){\line(0,1){1}}
\put(2,1.5){\line(1,0){1}}
\put(2,0.5){\line(1,1){1}}
\put(2,1.5){\line(1,-1){1}}
\put(3,0.5){\line(0,1){1}}
\put(3,0.5){\line(1,1){1}}
\put(3,0.5){\line(2,1){2}}
\put(4,1.5){\line(1,0){1}}
\put(5,0.5){\line(0,1){1}}
\put(5,0.5){\line(1,1){1}}
\put(2,0.1){\makebox(0,0){$v_{1}$}}
\put(3,0.1){\makebox(0,0){$v_{2}$}}
\put(2,1.85){\makebox(0,0){$v_{3}$}}
\put(3,1.85){\makebox(0,0){$v_{4}$}}
\put(4,1.85){\makebox(0,0){$v_{5}$}}
\put(5,0.1){\makebox(0,0){$v_{6}$}}
\put(5,1.85){\makebox(0,0){$v_{7}$}}
\put(6,1.85){\makebox(0,0){$v_{8}$}}
\put(6,0.1){\makebox(0,0){$v_{9}$}}
\put(1,1){\makebox(0,0){$G_{1}$}}
\multiput(9,0.5)(2,0){2}{\circle*{0.29}}
\multiput(12,0.5)(0,1){2}{\circle*{0.29}}
\multiput(9,1.5)(1,0){3}{\circle*{0.29}}
\put(9,0.5){\line(1,0){3}}
\put(9,1.5){\line(1,0){2}}
\put(9,0.5){\line(0,1){1}}
\put(9,0.5){\line(1,1){1}}
\put(9,0.5){\line(2,1){2}}
\put(9,1.5){\line(2,-1){2}}
\put(10,1.5){\line(1,-1){1}}
\put(11,0.5){\line(0,1){1}}
\put(12,0.5){\line(0,1){1}}
\put(9,0.1){\makebox(0,0){$u_{1}$}}
\put(11,0.1){\makebox(0,0){$u_{5}$}}
\put(10,1.85){\makebox(0,0){$u_{3}$}}
\put(11,1.85){\makebox(0,0){$u_{4}$}}
\put(9,1.85){\makebox(0,0){$u_{2}$}}
\put(12,0.1){\makebox(0,0){$u_{6}$}}
\put(12,1.85){\makebox(0,0){$u_{7}$}}
\put(8,1){\makebox(0,0){$G_{2}$}}
\end{picture}\caption{$G_{1},G_{2}$ are non-K\"{o}nig-Egerv\'{a}ry graphs.}%
\label{Fig5}%
\end{figure}

The graph $G_{1}$ from Figure \ref{Fig5} has $\alpha\left(  G_{1}\right)  =4$,
\textrm{corona}$\left(  G_{1}\right)  =\left\{  v_{1},v_{3},v_{4},v_{5}%
,v_{7},v_{8},v_{9}\right\}  $, \textrm{core}$\left(  G_{1}\right)
=\{v_{8},v_{9}\}$, and then $2\cdot\alpha(G_{1})=8<2+7=\left\vert
\mathrm{core}(G_{1})\right\vert +\left\vert \mathrm{corona}(G_{1})\right\vert
.$

It has been shown in \cite{LevMan2006} that
\[
\alpha(G)+\left\vert \cap\left\{  V-S:S\in\Omega(G)\right\}  \right\vert
=\mu\left(  G\right)  +\left\vert \mathrm{core}(G)\right\vert
\]
is satisfied by every K\"{o}nig-Egerv\'{a}ry graph\textit{ }$G$, and taking
into account that clearly
\[
\left\vert \cap\left\{  V-S:S\in\Omega(G)\right\}  \right\vert =\left\vert
V\left(  G\right)  \right\vert -\left\vert \cup\left\{  S:S\in\Omega
(G)\right\}  \right\vert ,
\]
we infer that the K\"{o}nig-Egerv\'{a}ry graphs enjoy the following nice property.

\begin{proposition}
\label{prop1}If $G$ is a \textit{K\"{o}nig-Egerv\'{a}ry graph}, then
\[
2\cdot\alpha(G)=\left\vert \mathrm{core}(G)\right\vert +\left\vert
\mathrm{corona}(G)\right\vert .
\]

\end{proposition}

It is worth mentioning that the converse of Proposition \ref{prop1} is not
true. For instance, see the graph $G_{2}$ from Figure \ref{Fig5}, which has
$\alpha\left(  G_{2}\right)  =3$, \textrm{corona}$\left(  G_{2}\right)
=\left\{  u_{2},u_{4},u_{6},u_{7}\right\}  $, \textrm{core}$\left(
G_{2}\right)  =\left\{  u_{2},u_{4}\right\}  $, and then $2\cdot
\alpha(G)=6=2+4=\left\vert \mathrm{core}(G_{2})\right\vert +\left\vert
\mathrm{corona}(G_{2})\right\vert .$

The \textit{vertex covering number} of $G$, denoted by $\tau(G)$, is the
number of vertices in a minimum vertex cover in $G$, that is, the size of any
smallest vertex cover in $G$. Thus we have $\alpha(G)+\tau(G)=\left\vert
V\left(  G\right)  \right\vert $. Since
\[
\left\vert V\left(  G\right)  \right\vert -\left\vert \cup\left\{
S:S\in\Omega(G)\right\}  \right\vert =\left\vert \cap\left\{  V-S:S\in
\Omega(G)\right\}  \right\vert ,
\]
Corollary \ref{cor2} implies the following.

\begin{corollary}
\cite{GitVal2006} If $G=\left(  V,E\right)  $, then $\alpha(G)-\left\vert
\mathrm{core}(G)\right\vert \leq\tau(G)-|\cap\left\{  V-S:S\in\Omega
(G)\right\}  |.$
\end{corollary}

Applying Matching Lemma \emph{(i)} to $\Lambda=\Omega(G)$ we immediately
obtain the following.

\begin{corollary}
\label{cor1}\cite{BorosGolLev2002} For every $S\in\Omega(G)$, there is a
matching from $S-\mathrm{core}(G)$ into $\mathrm{corona}(G)-S$.
\end{corollary}

Since every maximum clique of $G$ is a maximum independent set of
$\overline{G}$, Corollary \ref{cor3} is equivalent to the \textquotedblleft%
\textit{Clique Collection Lemma}\textquotedblright\ due to Hajnal.

\begin{corollary}
\label{cor4}\cite{Hajnal1965} If $\Gamma$ is a collection of maximum cliques
in $G$, then
\[
\left\vert \cap\Gamma\right\vert \geq2\cdot\omega(G)-\left\vert \cup
\Gamma\right\vert .
\]

\end{corollary}

Another appl\emph{ication of Matching Lem}ma is the \textquotedblleft%
\textit{Maximum Stable Set Lemma}\textquotedblright\ due to Berge.

\begin{corollary}
\cite{Berge1981}, \cite{Berge1985} An independent set $X$ is maximum if and
only if every independent set $S$ disjoint from $X$ can be matched into $X$.
\end{corollary}

\begin{proof}
Matching Lemma \emph{(ii)} is, essentially, the \textquotedblleft%
\textit{if}\textquotedblright\ part of corollary.

For the \textquotedblleft\textit{only if}\textquotedblright\ part we proceed
as follows. According to the hypothesis, there is a matching from $S-X=S-S\cap
X$ into $X$, in fact, into $X-S\cap X$, for each $S\in\Omega\left(  G\right)
-\left\{  X\right\}  $. Hence, we obtain
\[
\alpha\left(  G\right)  =\left\vert S\right\vert =\left\vert S-X\right\vert
+\left\vert S\cap X\right\vert \leq\left\vert X-S\cap X\right\vert +\left\vert
S\cap X\right\vert =\left\vert X\right\vert \leq\alpha\left(  G\right)  ,
\]
which clearly implies $X\in\Omega\left(  G\right)  $.
\end{proof}

\section{Conclusions}

In this paper we have proved the \textquotedblleft\textit{Set and Collection
Lemma}\textquotedblright, which has been crucial in order to obtain a number
of alternative proofs and/or strengthenings of some known results. Our main
motivation has been the \textquotedblleft\textit{Clique Collection
Lemma}\textquotedblright\ due to Hajnal \cite{Hajnal1965}. Not only this lemma
is beautiful but it is in continuous use as well. Let us only mention its two
recent applications in \cite{King2011, Rabern2011}.

Proposition \ref{prop1} claims that $2\cdot\alpha(G)=\left\vert \mathrm{core}%
(G)\right\vert +\left\vert \mathrm{corona}(G)\right\vert $ holds for every
K\"{o}nig-Egerv\'{a}ry graph\textit{ }$G$. Therefore, it is true for each very
well-covered graph $G$, \cite{LevMan1998}. Recall that $G$ is a \textit{very
well-covered} graph if $2\alpha(G)=\left\vert V\left(  G\right)  \right\vert
$, and all its maximal independent sets are of the same cardinality,
\cite{Favaron1982}. It is worth noting that there are other graphs enjoying
this equality, e.g., every graph $G$ having a unique maximum independent set,
because, in this case, $\alpha(G)=\left\vert \mathrm{core}(G)\right\vert
=\left\vert \mathrm{corona}(G)\right\vert $.

\begin{problem}
Characterize graphs satisfying $2\cdot\alpha(G)=\left\vert \mathrm{core}%
(G)\right\vert +\left\vert \mathrm{corona}(G)\right\vert $.
\end{problem}

Let us consider a dual problem. It is clear that for every graph $G$ there
exists a collection of maximum independent sets $\Lambda$ such that
$2\cdot\alpha(G)=\left\vert \cup\Lambda\right\vert +\left\vert \cap
\Lambda\right\vert $. Just take $\Lambda=\left\{  X\right\}  $ for some
maximum independent set $X$.

\begin{problem}
For a given graph $G$ find the cardinality of a largest collection of maximum
independent sets $\Lambda$ such that $2\cdot\alpha(G)=\left\vert \cup
\Lambda\right\vert +\left\vert \cap\Lambda\right\vert .$
\end{problem}

\section{Acknowledgments}

We express our gratitude to Pavel Dvorak for pointing out a gap in the proof
of Lemma \ref{MatchLem}.

\end{document}